\pdfoutput=1
\documentclass[11pt]{article}

\usepackage{graphicx}
\usepackage{amsmath}
\usepackage{dcolumn}
\usepackage{bm}
\usepackage{slashed}
\usepackage[utf8]{inputenc}
\usepackage{authblk}

\newcommand{\be}{\begin{equation}}
\newcommand{\ee}{\end{equation}}
\newcommand{\ba}{\begin{eqnarray}}
\newcommand{\ea}{\end{eqnarray}}
\newcommand{\no}{\nonumber \\}
\newcommand{\gsim}{\mathrel{\hbox{\rlap{\lower.55ex \hbox {$\sim$}}
                   \kern-.3em \raise.4ex \hbox{$>$}}}}
\newcommand{\lsim}{\mathrel{\hbox{\rlap{\lower.55ex \hbox {$\sim$}}
                   \kern-.3em \raise.4ex \hbox{$<$}}}}

\def\vk{{\vec k}}

\def\roughly#1{\mathrel{\raise.3ex\hbox{$#1$\kern-.75em%
\lower1ex\hbox{$\sim$}}}}
\def\lsim{\roughly<}
\def\gsim{\roughly>}

\def\({\left(}
\def\){\right)}
\def\[{\left[}
\def\]{\right]}
\def\<{\langle}
\def\>{\rangle}

\def\d{{\delta}}
\def\D{{\Delta}}
\def\o{{\omega}}
\def\O{{\Omega}}
\def\e{{\epsilon}}

\def\a{{\alpha}}
\def\b{{\beta}}
\def\c{{\chi}}

\def\g{{\gamma}}
\def\G{{\Gamma}}

\def\m{{\mu}}
\def\n{{\nu}}
\def\r{{\rho}}

\def\S{{\Sigma}}
\def\t{{\tau}}
\def\th{{\theta}}

\def\ps{{\psi}}

\newcommand{\tr}{\text{tr}}

\newcommand{\st}{{}^*}

\newcommand{\Dp}{\Delta_+}
\newcommand{\Dm}{\Delta_-}

\newcommand{\lag}{\langle}
\newcommand{\rag}{\rangle}

\title{\bf Fluctuation and Dissipation of Axial Charge from Massive Quarks}

\author[1]{De-fu Hou\thanks{houdf@mail.ccnu.edu.cn}}
\author[2]{Shu Lin\thanks{linshu8@mail.sysu.edu.cn}}
\affil[1]{Institute of Particle Physics (IOPP) and Key Laboratory of Quark and Lepton Physics (MOE),  Central China Normal University, Wuhan 430079, China}
\affil[2]{School of Physics and Astronomy, Sun Yat-Sen University, Zhuhai, 519082, China}

\date{\today}

\begin{document}

\maketitle

\begin{abstract}
In quantum chromodynamics (QCD), axial charge is known to be non-conserved due to chiral anomaly and non-vanishing quark mass. In this paper, we explore the role of quark mass in axial charge fluctuation and dissipation. We present two separate calculations of axial charge correlator, which describe dynamics of axial charge. The first is free quarks at finite temperature. We find that axial charge can be generated through effective quantum fluctuations in free theory. However the fluctuation does not follow a random walk behavior. Due to the presence of axial symmetry breaking mass term, the axial charge also does not settle asymptotically to the thermodynamic limit given by susceptibility. The second calculation is in weakly coupled quark gluon plasma (QGP). We find in the hard thermal loop (HTL) approximation, the quark-gluon interaction leads to random walk growth of axial charge, but dissipation is not visible. We estimate relaxation time scale for axial charge, finding it lies beyond the HTL regime.
\end{abstract}

\newpage

\section{Introduction}

The chiral anomaly is one of the most intriguing discovery in quantum field theory. Over the past ten years, its manifestations in macroscopic phenomena such as the chiral magnetic effect and chiral vortical effect have triggered significant interests across different communities \cite{Vilenkin:1980fu,Kharzeev:2004ey,Kharzeev:2007tn,Fukushima:2008xe,Fukushima:2010vw,Son:2012bg}. There have been continuous efforts in searching for CME and CVE in quark gluon plasma produced in heavy ion collisions \cite{Adamczyk:2014mzf,Abelev:2012pa,Sirunyan:2017quh}, as well as in Weyl semi-metal \cite{Li:2014bha,Gooth:2017mbd}. For comprehensive reviews of current status, we refer to \cite{Kharzeev:2015znc,Liao:2014ava,Huang:2015oca}. Effective description of CME and CVE have been developed for chiral fermions, including anomalous hydrodynamics \cite{Son:2009tf,Neiman:2010zi,Landsteiner:2011cp}, chiral kinetic theory \cite{Son:2012wh,Son:2012zy,Stephanov:2012ki,Pu:2010as,Chen:2012ca,Mueller:2017lzw,Mueller:2017arw} and holography \cite{Yee:2009vw,Landsteiner:2011iq,Hoyos:2011us,Lin:2013sga,Jimenez-Alba:2014iia,Jimenez-Alba:2015awa}. 
Both theoretical frameworks reveal beautiful structures in the chiral limit. In most phenomenological applications of the two frameworks, axial charge density is needed as an input, usually modeled by axial charge chemical potential $\m_5$. Possible issues with using spacetime dependent $\m_5$ is pointed out by one of the authors \cite{Wu:2016dam}. To use $\m_5$ properly, a better understanding of the dynamics of axial charge is needed.

One of the well-known generation mechanism of axial charge is through topological fluctuation of gluon field \cite{Moore:2010jd,Akamatsu:2015kau}. Due to fluctuation-dissipation theorem, this mechanism can also cause damping of axial charge. Most phenomenological studies ignore the damping effect in the dynamics of axial charge. The interplay of generation and damping is known to lead to interesting dynamics of axial charge \cite{Iatrakis:2015fma}. In addition to topological fluctuation of gluons, fermion mass violates axial charge conservation explicitly. Questions on the role of fermion mass also arises in different context. On the theoretical side, any fundamental fermion is known to carry mass. Knowing how fermion mass modifies the existing frameworks is a key ingredient. On the phenomenological side, quantifying the magnitude of the mass effect is in need for reliable modeling. The damping effect is first discussed in \cite{Grabowska:2014efa} for electron in neutron star. The generation effect is proposed by one of the authors \cite{Guo:2016nnq,Guo:2016dnm} for supersymmetric gauge theory based on a holographic model.

This paper aims at providing a unified description of the two effects in the same setting. To be specific, we study dynamics of total axial charge in QCD in the weakly coupled regime, where perturbative calculation is possible. The results apply equally well to QED. For pedagogical reasons, we begin with axial charge dynamics in free quark theory in Section II. We then move on to carry out the same study in weakly coupled QCD plasma in Section III. We find that unlike free theory, weakly coupled plasma can generate axial charge through quark mass term similar to the topological fluctuation of gluons. The rate of generation, to be coined mass diffusion rate, is numerically much smaller than the topological fluctuation rate at physically relevant coupling for strange quark mass. We summarize in Section IV.

\section{Axial Charge Dynamics in Free Theory}
For simplicity, we first study axial charge dynamics in free theory. This allows us to set the stage and gain insights into the dynamics. We first note that classical fermions satisfying Dirac equation do not have a net axial charge, even though axial symmetry is broken by fermion mass. In order to generate net axial charge, we need quantum fluctuation to push fermions off-shell. The quantity characterizing axial charge dynamics is the following Wightman correlator
\begin{align}\label{wightman}
\int dtd^3{\bf x}e^{iq_0 t}\lag \ps^+\g^5\ps(t,x)\;\ps^+\g^5\ps\big(0)\rag.
\end{align}
This describes dynamics of total axial charge $N_5=\int d^3{\bf x}\ps^+\g^5\ps(x)$.
Contribution to \eqref{wightman} comes from a simple quark loop diagram. The contribution is given by
\begin{align}\label{pi12}
G^>(Q)=\int\frac{d^4K}{(2\pi)^4}\text{Tr}S_{21}(K)S_{12}(K-Q),
\end{align}
with $K=(k_0,\,\vk)$ and $Q=(q_0,\,0)$.
Let us first consider the case $q_0>0$. Using delta functions in fermion propagators $S_{21}$ and $S_{12}$, we find the only possible kinematics is $k_0=E_k$, and $k_0-q_0=-E_k$, with $E_k=\sqrt{k^2+m^2}$. It is not difficult to evaluate the integral to obtain:
\begin{align}\label{pi12_r}
&G^>(q_0)=\frac{N_fN_c}{\pi}\tilde{f}\(-\frac{q_0}{2}\)^2\sqrt{\(\frac{q_0}{2}\)^2-m^2}\frac{2m^2}{q_0}\th(q_0-2m),\quad q_0>0,\\
&\text{with}\;
\tilde{f}(-q_0/2)=\frac{1}{e^{-\b q_0/2}+1}. \nonumber
\end{align}
This gives the spectral function $\r(q_0)$
\begin{align}
&\r(q_0)=\frac{G^>(q_0)}{1+f(q_0)}=\frac{N_fN_c}{\pi}\frac{\tilde{f}(q_0/2)}{f(q_0/2)}\sqrt{\(\frac{q_0}{2}\)^2-m^2}\frac{2m^2}{q_0}\th(q_0-2m),\quad q_0>0,\\
&\text{with}\;
{f}(q_0/2)=\frac{1}{e^{\b q_0/2}-1}. \nonumber
\end{align}
For the case $q_0<0$, we use the representation $G^>(q_0)=\r(q_0)(1+f(q_0))$ and the property $\r(-q_0)=-\r(q_0)$ to obtain
\begin{align}\label{pi21_r}
G^>(q_0)=-\frac{N_fN_c}{\pi}\tilde{f}\(-\frac{q_0}{2}\)^2\sqrt{\(\frac{q_0}{2}\)^2-m^2}\frac{2m^2}{q_0}\th(-q_0-2m), \quad q_0<0.
\end{align}
The above evaluation misses a contribution proportional to $\d(q_0)$. It is known that such a contribution corresponds to susceptibility for conserved charge \cite{McLerran:1987pz}. We can include this contribution by assuming the following decomposition of $G^>$
\begin{align}\label{decomp}
G^>=G(q_0,T,m)+\d(q_0)F(T,m).
\end{align}
We have already obtained $G(q_0,T,m)$ in \eqref{pi12_r} and \eqref{pi21_r}. $F(T,m)$ can be obtained by
\begin{align}\label{F}
F(T,m)=\lim_{\e\to0}\int_{-\e}^{\e}dq_0G^>(q_0,T,m)
\end{align}
Using \eqref{pi12} for the evaluation of \eqref{F}, we obtain
\begin{align}\label{F_int}
F(T,m)=\int_0^\infty dk\frac{4}{\pi}\frac{k^4}{E_k^2}\tilde{f}(E_k)(1+\tilde{f}(E_k)).
\end{align}

\begin{figure}
\includegraphics[width=0.5\textwidth]{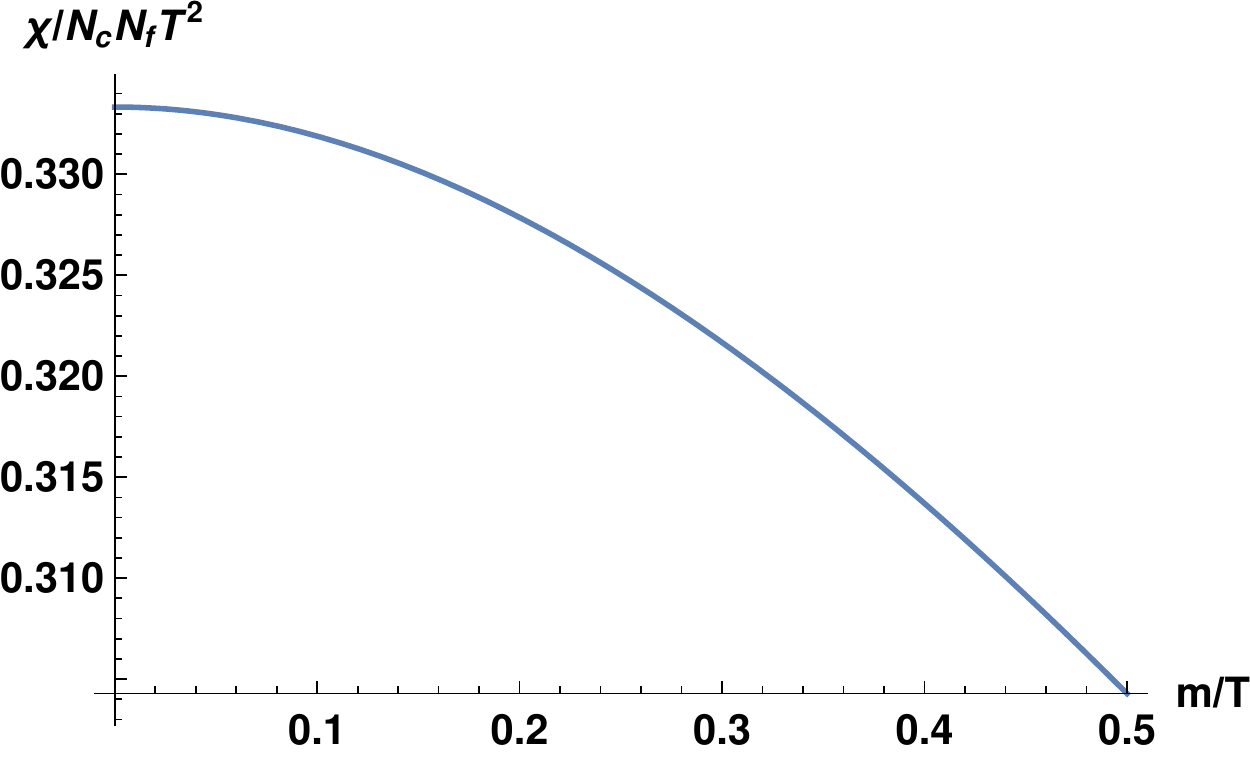}
\caption{\label{suscept_fig}Normalized susceptibility versus $m/T$ for both $N_5$ and $N$. It reduces to known result $\c=N_cN_fT^2/3$ in the massless limit.}
\end{figure}

It is instructive to compare \eqref{decomp} with Wightman correlator for $N=\int d^3{\bf x}\psi^+\psi(x)$. Following similar procedure, we would obtain the same expression \eqref{decomp} but without $G(q_0,T,m)$. It is known that for conserved charge $N$, $G^>=2\pi T\d(q_0)\c$, thus $F(T,m)$ is simply related to susceptibility $F(T,m)=2\pi T\c$.
By analogy, we define susceptibility of $N_5$ from contribution of $F$ by $\c=F/(2\pi T)$. We plot the $m$-dependence of $\c$ in Figure.\ref{suscept_fig}.

The fluctuation of $N_5$ is characterized by the correlator $\lag\D N_5(t)^2\rag\equiv\lag(N_5(t)-N_5(0))^2\rag$, which can be expressed by $G^>$ as
\begin{align}\label{N5_fluc}
\lag\D N_5(t)^2\rag=V\int \frac{dq_0}{2\pi}(2-e^{iq_0t}-e^{-iq_0t})G^>(q_0),
\end{align}
with $V=\int d^3x$ being the volume factor. Let us look at the contribution from $F$ and $G$ separately. The evaluation of the former is subtle: a naive integration of $\d(q_0)$ gives a vanishing result. However, on general ground we expect as $t\to\infty$,
\begin{align}\label{suscept}
\lag\D N_5(t)^2\rag&=\lag N_5(t)^2\rag+\lag N_5(0)^2\rag-\lag N_5(t)N_5(0)\rag-\lag N_5(0)N_5(t)\rag
\to 2\lag N_5(0)^2\rag,
\end{align}
where we used $\lag N_5(t)N_5(0)\rag\to0$ and $\lag N_5(t)^2\rag=\lag N_5(0)^2\rag$\footnote{In \cite{Iatrakis:2015fma}, the same quantity is calculated in the stochastic hydrodynamics framework. $\c TV$ is obtained instead. The reason is we set initial $N_5(0)=0$. This amounts to subtracting a baseline for the fluctuation.}. Taking contribution to $\lag N_5(0)^2\rag$ from $F$, we would instead obtain:
\begin{align}
\lag\D N_5(t)^2\rag_F=2\c TV.
\end{align}
The origin of the disagreement is that the two limits $q_0\to 0$(or $t\to\infty$) and $k\to 0$(or $V\to\infty$) do not commute. In \eqref{N5_fluc}, we take $V\to\infty$ first while in \eqref{suscept}, we take $t\to\infty$ first. Physically they are not equivalent: since conserved charge can fluctuate only through charge exchange with heat bath, taking $V\to\infty$ requires larger and larger heat bath, and consequently longer and longer equilibration time. To reproduce \eqref{suscept}, we should take the limit $t\to\infty$ first, which amounts to dropping the rapid oscillating terms in \eqref{N5_fluc}. The resulting $\lag\D N_5(t)^2\rag$ indeed obtain \eqref{suscept}.

Now we turn to evaluation of the contribution from $G$. This is intrinsic to breaking of the axial symmetry. It corresponds to fluctuation by itself, not relying on charge exchange with heat bath. It is easy to see from \eqref{pi12_r} that the fluctuation also exist in vacuum.
Plugging \eqref{pi12_r} and \eqref{pi21_r} into \eqref{N5_fluc}, we find
the integral contains a UV divergence. We regularize by subtracting the vacuum contribution:
\begin{align}\label{N5_free}
\lag\D N_5(t)^2\rag_G=VN_fN_c\int_0^\infty \frac{dq_0}{2\pi}\frac{2(1-\cos(q_0t))}{\pi\hbar^3}\big[\frac{\tilde{f}(q_0/2)^2}{\tilde{f}(q_0)}-1\big]\sqrt{\(\frac{q_0}{2}\)^2-m^2}\frac{m^2}{q_0}\th(q_0-2m).
\end{align}
We have restored factor of $\hbar$ in \eqref{N5_free}. Note that on the left hand side (LHS), $N_5$ is dimensionless. On the right hand side (RHS), the dimension reads $(\text{energy})^3(\text{length})^3/\hbar^3$, also dimensionless. 
We point out two counter-intuitive features of \eqref{N5_free}. The fluctuation of $N_5$ contains explicit factor of $\hbar$, indicating it is a consequence of quantum fluctuation. However, we know in free quark case there is no interaction to induce quantum fluctuation. The other odd feature is that the regularized fluctuation is negative (as is clear from the negativity of the square bracket)! It means that the fluctuation at finite temperature is smaller compared to that in vacuum.

The two seemingly odd features are in fact related: Although quarks are free at Lagrangian level, Fermi-Dirac statistics obeyed by quarks in equilibrium provides effective interaction, thus quantum fluctuation is present. Furthermore, this also gives a quantitatively explanation of the negative sign in the regularized fluctuations. The effect of Fermi-Dirac statistics becomes prominent as we lower the temperature. In the vacuum case, quantum fluctuation is maximal, thus the vacuum fluctuation is larger than any finite temperature fluctuation, giving rise to negative regularized fluctuation.
\begin{figure}
\includegraphics[width=0.5\textwidth]{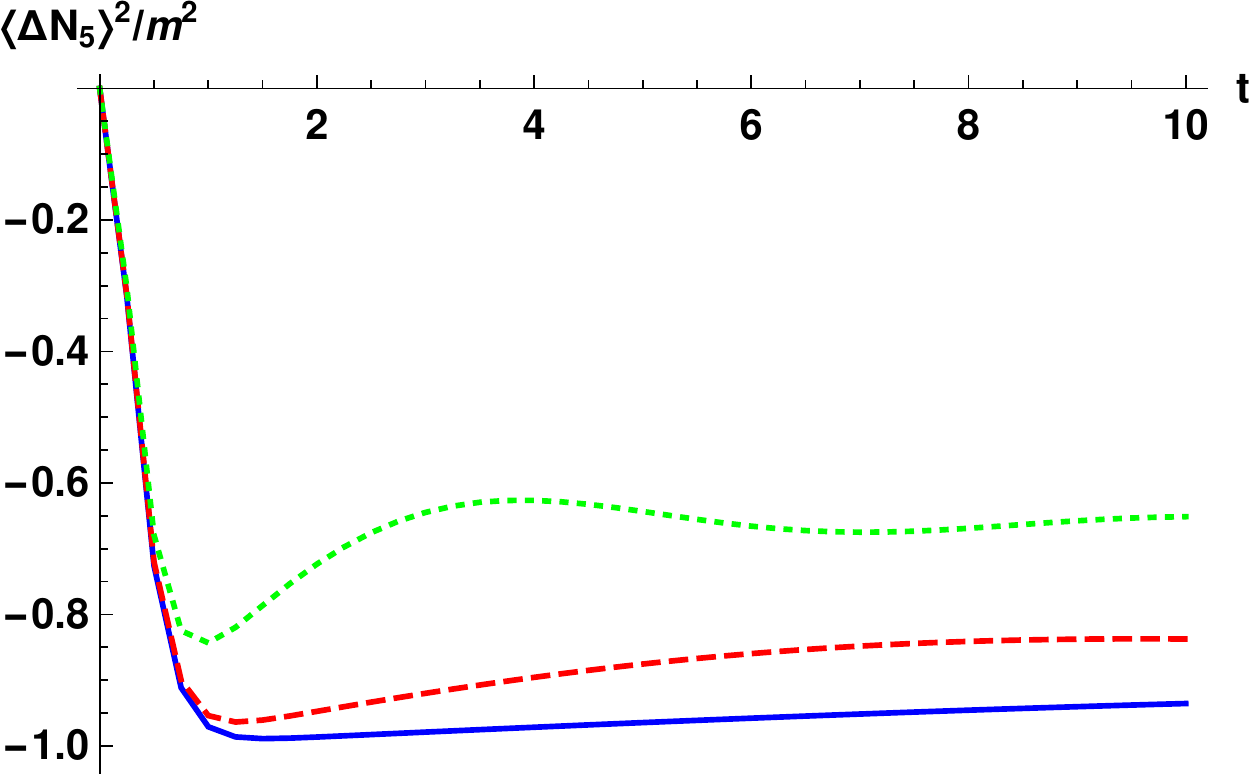}
\caption{\label{N5_fig}Contribution from intrinsic fluctuation $\lag\D N_5(t)^2\rag/m^2$ for different masses: blue solid for $m=1/10$, red dashed for $m=1/5$ and green dotted for $m=1/2$. The unit is set by $T=1$. The fluctuation is characterized by an initial rise followed by oscillatory decay to asymptotic value. The case with larger mass shows more rapid convergence to asymptotic value.}
\end{figure}
\eqref{N5_free} can be evaluated numerically. We include the time evolution of \eqref{N5_free} for different $m$ in Figure.~\ref{N5_fig}. The fluctuation is characterized by an initial rise followed by oscillatory decay to asymptotic value. Fig.~\ref{N5_fig} suggests the initial rise satisfies the scaling $\lag\D N_5(t)^2\rag\propto m^2$. Since the mass term is the source of fluctuation, the $m^2$ dependence as lowest order expansion is expected from analyticty in $m$. Non-analyticity can occur in the presence of external field due to Schwinger effect \cite{Fukushima:2010vw,Copinger:2018ftr}. Furthermore, if we regard the period of oscillation as relaxation time, Fig.~\ref{N5_fig} also implies shorter relaxation time at larger mass, which is consistent with expectation on general grounds.

To summarize, we find the fluctuation of $N_5$ contains two contributions \eqref{suscept} and \eqref{N5_free}:
\begin{align}\label{N5_sum}
\lag\D N_5(t)^2\rag&=2V\c T+V\int\frac{dq_0}{2\pi}(2-e^{iq_0t}-e^{-iq_0t})G(q_0),
\end{align}
where in the second line the limit $t\to\infty$ should be taken.
The $\c$ term arises from charge exchange with heat bath. The term proportional to $G$ is intrinsic to breaking of the axial symmetry. It exists without a heat bath. We could have view the second term as correction to susceptibility. But this interpretation is misleading. Note that the second term is not necessarily proportional to temperature as it arises from quantum fluctuation.
In the next section, we will focus on the modification of the intrinsic fluctuation by interaction.
%
%

\section{Axial Charge Dynamics in Weakly Coupled QGP}
After the warm-up, we move on to the calculation in weakly coupled QGP. We expect the same structure of the Wightman correlator as \eqref{decomp}. The term proportional to $F$ is related to susceptibility of $N$. It has been calculated in perturbation theory \cite{Blaizot:2002xz}. The other term is entirely due to non-conservation of $N_5$. 
%
We will calculate $G^>_G$ in weakly coupled QGP, with the subscript indicating it only contains the $G$ term. We start with retarded correlator, whose imaginary part is related to the Wightman correlator.
\begin{align}\label{GR_def}
G^R(q_0)\equiv\int dtd^3{\bf x}e^{iq_0 t}\lag \big[\ps^+\g^5\ps(x),\;\ps^+\g^5\ps\big(0)]\rag.
\end{align}
We will proceed in imaginary time formalism and analytically continue to real time in the end. We work in the HTL approximation at one loop order. It is known from the calculation of susceptibility that one loop result of HTL is not complete (\cite{Blaizot:2002xz} and references therein). However the main purpose of this paper is to demonstrate the diffusive behavior of axial charge from quark mass effect, we restrict ourselves to one loop order, and leave more refined studies for future work.

\begin{figure}
\includegraphics[width=0.3\textwidth]{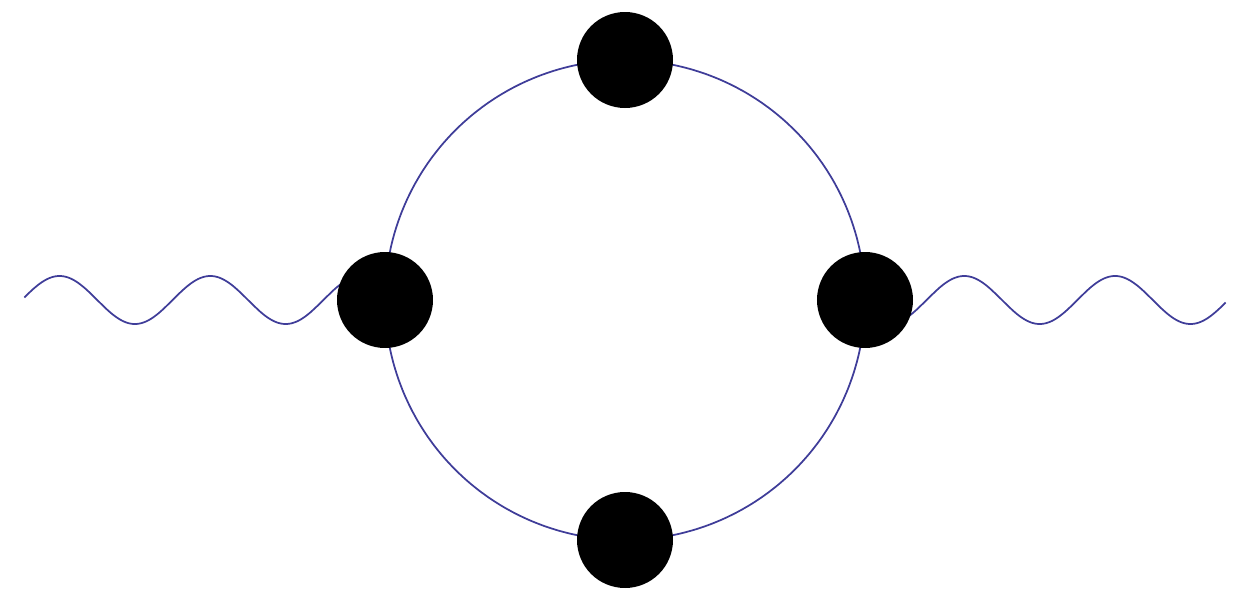}
\includegraphics[width=0.3\textwidth]{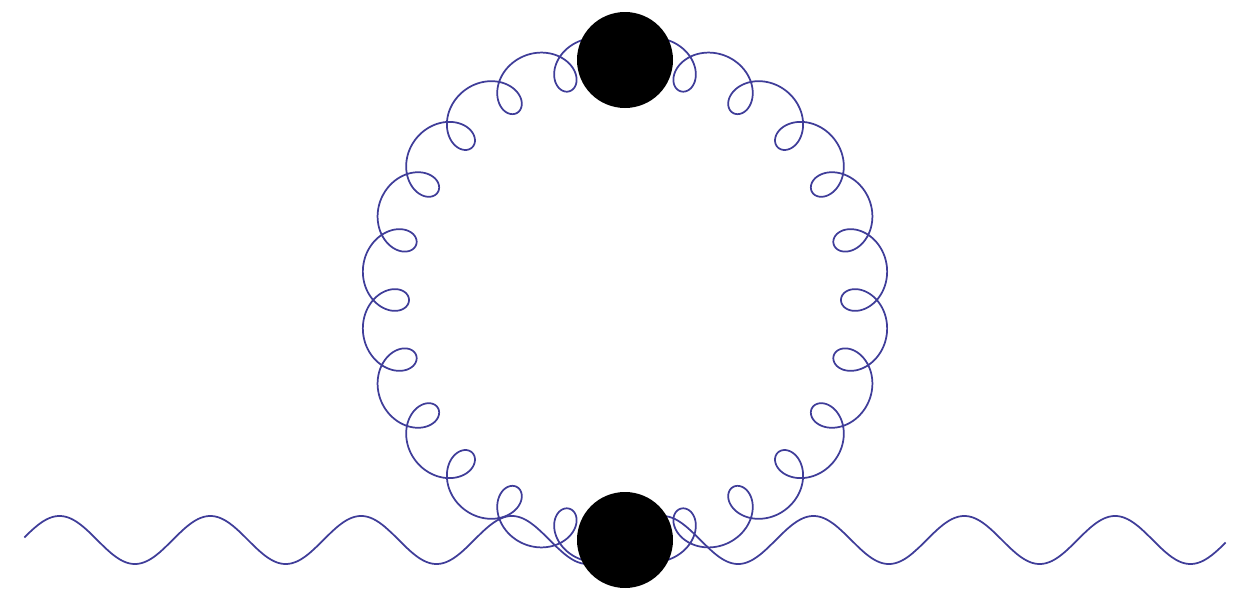}
\includegraphics[width=0.3\textwidth]{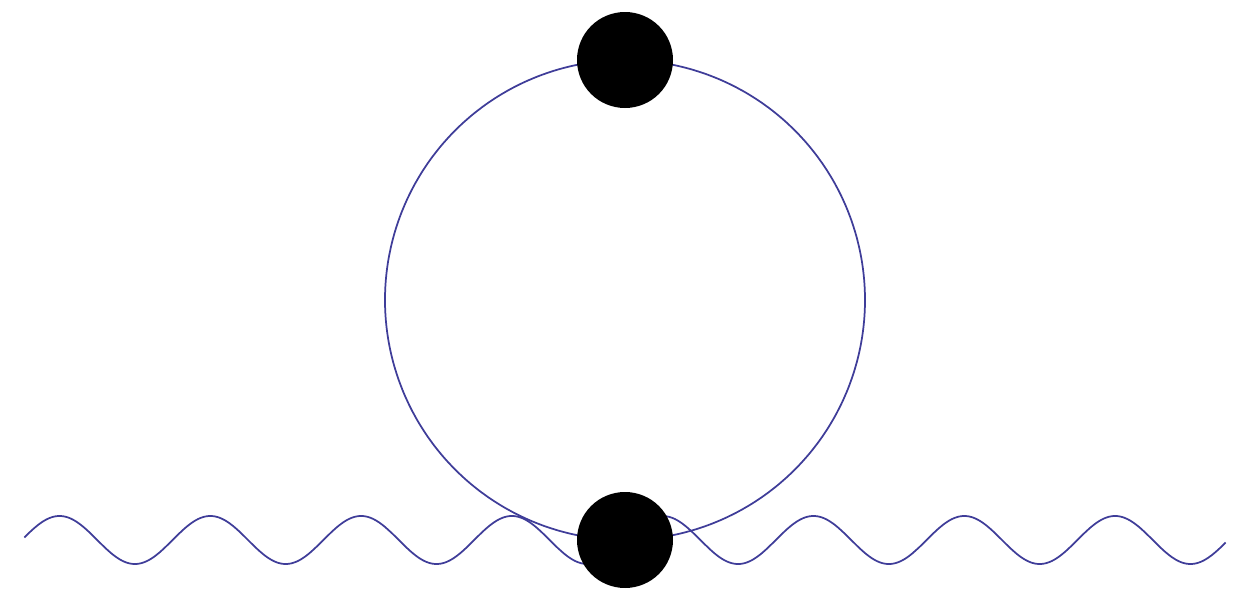}
\caption{\label{feyn_htl}Three leading one loop diagrams contributing to \eqref{GR_def} in the HTL approximation.}
\end{figure}

At one loop order, $N_5$ correlator receives contributions from three diagrams as shown in Figure.~\ref{feyn_htl}. The first diagram contains a soft quark loop with pseudophoton-quark vertices ($\tilde\g qq$). The second diagram contains a soft gluon loop with two-pseudophoton-two-gluon vertex ($2\tilde\g2g$) and the third diagram contains a quark loop with a two-pseudophoton-two-quark vertex ($2\tilde\g2q$) vertex. Here we used $\tilde\g$ to denote pseudophoton leg. These resummed vertices are to be evaluated for HTL diagrams. When the quark mass $m=0$, we can easily show by commuting $\g^5$ with $\g^\m$ the following relations
\begin{align}\label{vertices}
&\tilde\g qq=\g qq\times \g^5, \no
&2\tilde\g2g=2\g2g, \no
&2\tilde\g 2q=2\g2q
\end{align}
We have used schematic notations. The first line of \eqref{vertices} means pseudophoton-quark vertex equals photon-quark vertex times $\g^5$ and similarly for the second and third equalities. When $m\ne0$, in general all the vertices involving pseudophoton receive corrections from $m$. To simplify the computation, we take quark mass to be soft, i.e. $m\sim gT$. We stress that this is ``current quark mass'', not to be confused with thermal quark mass, which will appear below as $m_f$. The current quark mass itself should be $T$ independent. The relation $m\sim gT$ is only meant for numerical values for specific $m$, $g$ and $T$.

In the spirit of HTL, we will drop any contributions at $O(\frac{m^2}{T^2})$. This allows \eqref{vertices} to hold in this approximation. We are ready to write down explicit expressions of the vertices involved
\begin{align}\label{vertices2}
&\st\G_{5\m}(P_1,P_2)=\(\g_\m-m_f^2\int\frac{d\O}{4\pi}\frac{\hat{K}_\m\hat{\slashed{K}}}{(P_1\cdot\hat{K})(P_2\cdot\hat{K})}\)\g_5, \\
&\st\G_{\m\n}(P_1,P_2,Q_1,Q_2)=-m_f^2\int\frac{d\O}{4\pi}\frac{\hat{K}_\m\hat{K}_\n\hat{\slashed{K}}}{(P_1+Q_1)\cdot\hat{\slashed{K}}(P_2-Q_1)\cdot\hat{\slashed{K}}}\bigg[\frac{1}{P_1\cdot\hat{K}}+\frac{1}{P_2\cdot\hat{K}}\bigg],
\end{align}
with $P_1$ and $P_2$ being quark momenta and $Q_1$ being one of the gluon momenta.
$m_f^2=\frac{1}{8}C_Fg^2T^2$ is thermal quark mass, not to be confused with current quark mass $m$. The null vector is defined as $\hat{K}=(-i,\;\hat{k})$. The remaining $2\tilde{\g}2g$ vertex is obtainable by sending two generators to $1$ in four-gluon vertex. This leads to a vanishing result, thus the second diagram drops out.
Unlike vertices, the resummed quark propagator does get modification due to quark mass as follows:
\begin{align}\label{propagator}
\st S(P)=\frac{1}{\slashed{P}+\S+m}=\frac{\frac{1}{2}(\Dp+\Dm)i\g_4+\frac{1}{2}(\Dp-\Dm)-m\Dp\Dm}{1-m^2\Dp\Dm},
\end{align}
with $\g_p={\hat p}\cdot{\vec\g}$ and
\begin{align}\label{deltas}
&\S=\frac{m_f^2}{p}\big[i\g_4 Q_0\(\frac{\o}{p}\)+\g_p\(1-\frac{i\o}{p}Q_0\(\frac{\o}{p}\)\)\big] \no
&\frac{1}{\Delta_\pm(P)}=i\o\mp p-\frac{m_f^2}{p}\big[ Q_0\(\frac{i\o}{p}\)\mp Q_1\(\frac{i\o}{p}\) \big].
\end{align}

Our calculation heavily relies on Ward identities. We note the $\g qq$ vertex and $2\g2q$ vertex satisfy the following Ward identities:
\begin{align}\label{Ward}
&Q_{1\m}\st\G_{\m\n}(P_1,P_2,Q_1)=\G_\n(P_1,P_2-Q_1)-\G_\n(P_1+Q_1,P_2), \no
&(P_1-P_3)_\m\st\G_{\m}(P_1,P_3)=\S(P_1)-\S(P_3).
\end{align}
For our purpose, we take quark momenta as $P_1=P_2=P$, $P_3=P'$ and pseudophoton momentum $Q=(-\varpi,{\bf 0})=P-P'$. Consequently $\st\G_{44}$ and $\st\G_4$ can be uniquely fixed by Ward identities as
\begin{align}\label{Ward_rep}
&-\varpi\st\G_{44}(P,P,Q)=\st\G_4(P,P-Q)-\st\G_4(P+Q,P), \no
&-\varpi\st\G_4(P,P')=\S(P)-\S(P').
\end{align}

We proceed by evaluating the tadpole diagram
\begin{align}\label{3rd}
\int\frac{d^4p}{(2\pi)^4}\tr\st\G_{44}\st S(P)(-1),
\end{align}
The trace can be evaluated using \eqref{Ward_rep}
\begin{align}\label{tadpole}
&\int\frac{d^4p}{(2\pi)^4}\tr\st S(P)\big[\G_4(P,P')-\G_4(P+Q,P)\big]\frac{1}{\varpi} \no
=&\int\frac{d^4p}{(2\pi)^4}\big[\tr\st S(P)\st\G_4(P,P')-\tr\st S(P')\st\G_4(P,P')\big]\frac{1}{\varpi},
\end{align}
where in the second line, we make a change of variable: $P\to P'$ to the second $\G_4$.
This expression will be canceled by part of terms in quark-antiquark diagram. We proceed by simplifying the quark-antiquark diagram using \eqref{Ward_rep}. Note that $\st\G_4^5=\st\G_4\g^5$ and also $\st\G_4(P,P')=\st\G_4(P',P)$, which is obvious from \eqref{vertices2}, we have
\begin{align}\label{1st}
&\int\frac{d^4P}{(2\pi)^4}\tr\st S(P)\st\G_4^5(P,P')\st S(P')\st\G_4^5(P',P)(-1). \no
=&\int\frac{d^4P}{(2\pi)^4}\tr\frac{1}{{\slashed P}+\S+m}\st\G_4\g^5\frac{1}{{\slashed P'}+\S'+m}(\S-\S')\g^5\frac{1}{\varpi},
\end{align}
where we use short hand notation $\S=\S(P)$, $\S'=\S(P')$ and suppressed the argument of $\st\G_4$ for notational simplicity. Commuting $\g^5$ through only switch sign of mass in the second propagator. We can further simplify the expression by splitting $\S-\S'=({\slashed P}+\S+m)-({\slashed P}'+\S'-m)-({\slashed P}-{\slashed P}'+2m)$ and using cyclic property of trace to obtain
\begin{align}\label{qqbar}
&\int\frac{d^4P}{(2\pi)^4}\tr\frac{1}{{\slashed P}+\S+m}\st\G_4\frac{1}{{\slashed P'}+\S'-m}(\S-\S')\frac{1}{\varpi} \no
=&\int\frac{d^4P}{(2\pi)^4}\big[\tr\frac{1}{{\slashed P}'+\S'-m}\st\G_4-\frac{1}{{\slashed P}+\S+m}\st\G_4 
-\tr\frac{1}{{\slashed P}+\S+m}\st\G_4\frac{1}{{\slashed P}'+\S'-m}\times \no
&({\slashed P}-{\slashed P}'+2m)\big]\frac{1}{\varpi}.
\end{align}
The retarded correlator is given by sum of \eqref{tadpole} and \eqref{qqbar}.

It is instructive to look at the result in the massless limit first. Setting $m=0$, we immediately see the first two terms of \eqref{qqbar} cancel \eqref{tadpole} entirely, leaving only the third term of \eqref{qqbar}.
To evaluate the third term, we use \eqref{propagator} and the following explicit expression of $\G_4$.
\begin{align}
\st\G_4(P,P')=\big[\(1-\frac{m_f^2}{i{\varpi}p}\d Q_0(p,p')\)\g_4+\frac{m_f^2}{{\varpi}}\d Q_1(p,p')\g_p\big],
\end{align}
with $\d Q_n(P,P')=Q_n\(\frac{i\o}{p}\)-Q_n\(\frac{i\o'}{p}\)$. We can adopt a representation of $\G_4$ in terms of $\D_\pm\equiv\D_\pm(P)$ and $\D_\pm'\equiv\D_\pm(P')$ by using \eqref{deltas}.
\begin{align}
\st\G_4(P,P')=-\frac{1/\D_++1/\D_--1/\D_+'-1/\D_-'}{2\varpi}i\g_4+\frac{1/\D_+-1/\D_--1/\D_+'+1/\D_-'}{2\varpi}\g_p.
\end{align}
Taking the trace, we obtain
\begin{align}\label{massless}
\int\frac{d^4P}{(2\pi)^4}\frac{2i}{\varpi}(\D_-+\D_+-\D_-'-\D_+'),
\end{align}
which vanishes identically upon change of variable. This indicates that indeed the contribution we are after is intrinsic to breaking of the axial symmetry.

Now we move on to massive case. We note that the first two terms of \eqref{qqbar} combine with \eqref{tadpole} to give:
\begin{align}
\int\frac{d^4P}{(2\pi)^4}\tr\st\G_4\(\frac{1}{{\slashed P}'+\S'-m}-\frac{1}{{\slashed P}'+\S'+m}\)\frac{1}{\varpi},
\end{align}
which still vanishes upon taking the trace. The remaining terms are
\begin{align}
\int\frac{d^4P}{(2\pi)^4}(-)\tr\frac{1}{{\slashed P}+\S+m}\st\G_4\frac{1}{{\slashed P'}+\S'-m}\({\slashed P}-{\slashed P}'+2m\)\frac{1}{\varpi}.
\end{align}
We aim at calculating lowest order mass correction, which begins at order $O(m^2)$. It arises from expansion of denominator and numerator of propagators and mass term in the $\G_4$. It is easy to see that the expansion of the denominator still gives a vanishing result due to similar cancellation as the $O(m^0)$ result. The remaining correction can be organized as follows
\begin{align}\label{mass_corr}
\int\frac{d^4P}{(2\pi)^4}\frac{8m^2}{\varpi^2}\(-\D_+\D_--\D_+'\D_-'+\D_+\D_-'+\D_+'\D_-\).
\end{align}
We use the following formula to perform the frequency sum:
\begin{align}\label{freq_sum}
\text{Im}T\S_ng_1(i\o_n)g_2(i(\o_n-\o))&=\pi\(1-e^{\b q_0}\)\times \no
&\int_{-\infty}^{+\infty}\frac{dp_0}{2\pi}\frac{dp_0'}{2\pi}\tilde{f}(p_0)\tilde{f}(p_0')\d(q_0-p_0-p_0')\r_1(p_0)\r_2(-p_0').
\end{align}
Here $g_1$ and $g_2$ are two generic functions. $\r_1$ and $\r_2$ correspond to their spectral densities, $\r_1=-2Im g_1$, $\r_2=-2Im g_2$.
Note that the analytic continuation $i\o\to q_0+i\eta$ is taken after the frequency sum and only the imaginary part of the result is kept in \eqref{freq_sum}.
The frequency sum of the first two terms in \eqref{mass_corr} gives a contribution with $q_0=0$, which vanishes identically due to the factor $1-e^{\b q_0}$. The frequency sum of the remaining terms is given by
\begin{align}\label{mass_spec}
&\int\frac{d^3p}{(2\pi)^3}\int\frac{dp_0}{2\pi}\frac{dp_0'}{2\pi}\pi\(1-e^{\b q_0}\){\tilde f}(p_0){\tilde f}(p_0')\d(q_0-p_0-p_0')\frac{32m^2}{q_0^2}\times \no
&\(Im \D_+(p_0)Im\D_-(-p_0')+Im \D_-(p_0)Im\D_+(-p_0')\) \no
=&\int\frac{d^3p}{(2\pi)^3}\int\frac{dp_0}{2\pi}\frac{dp_0'}{2\pi}\pi\(1-e^{\b q_0}\){\tilde f}(p_0){\tilde f}(p_0')\d(q_0-p_0-p_0')\frac{32m^2}{q_0^2}\times \no
&\(Im \D_+(p_0)Im\D_+(p_0')+Im \D_-(p_0)Im\D_-(p_0')\).
\end{align}
We have used the property $Im\D_\pm(-p_0')=Im\D_\mp(p_0')$ in the second line.
%
Note that Wick rotation applies $N_5^4\to iN_5$, which gives an overall minus sign between correlator of $N_5^4$ and correlator of $N_5$.
Using KMS relation, we readily obtain
\begin{align}\label{G21_exp}
G^>_G(q_0)=&\int\frac{d^3p}{(2\pi)^2}\int\frac{dp_0}{2\pi}\frac{dp_0'}{2\pi}e^{\b q_0}{\tilde f}(p_0){\tilde f}(p_0')\d(q_0-p_0-p_0')\frac{32m^2}{q_0^2}\times \no
&\(Im \D_+(p_0)Im\D_+'(p_0')+Im \D_-(p_0)Im\D_-'(p_0')\).
\end{align}
Note that we have identified the contribution with $G^>_G$ because it arises entirely from quark mass breaking of the axial symmetry.
The spectral density appearing in \eqref{G21_exp} contains contribution from poles and a cut. The convolution of two spectral densities gives rise to contributions from the following types: pole-pole, pole-cut and cut-cut. Similar situation is encountered in the computation of soft dilepton production, showing a remarkable structure \cite{Braaten:1990wp,Moore:2006qn}.

For the purpose of demonstrating late time dynamics of $N_5$, we focus on the small $q_0$ regime.
In the limit $q_0\to 0$, we obtain $G^>_G\to\frac{4\G_m}{q_0^2}$, with $\G_m$ defined as
\begin{align}\label{Gm}
&\G_m=\int\frac{d^3p}{(2\pi)^2}\frac{dp_0}{2\pi}\frac{dp_0'}{2\pi}8m^2\d(p_0+p_0')\tilde{f}(p_0)\tilde{f}(p_0') \times \no
&\big[Im\Dp(p_0)Im\Dp(p_0')+Im\Dm(p_0)Im\Dm(p_0')\big].
\end{align}
$\G_m$ characterizes the rate of fluctuation of the axial charge.
To see this, we do the Fourier transform as follows
\begin{align}
\int d^3x\lag(n_5(t,x)-n_5(0))^2\rag&=\int\frac{dq_o}{2\pi} (2-e^{-iq_0t}-e^{iq_0t})G^>(q_0) \no
&\simeq\int\frac{dq_0}{2\pi}\frac{8\G_m}{q_0^2}(1-\cos q_0t)=4\G_m t.
\end{align}
This is the random walk growth of axial charge, with the growth rate given by \eqref{Gm}.

To evaluate $\G_m$, we note that the delta function constraint only allows for cut-cut contribution in the product $Im\D_\pm Im\D_\pm'$: The pole-pole contribution is excluded in the limit $q_0\to0$. The pole-cut contribution is possible only at large $p$, which is exponentially suppressed by the Fermi-Dirac distribution. The cut-cut contribution is not suppressed.
We send $\tilde{f}(p_0)\to 1/2$, $\tilde{f}(p_0')\to 1/2$ in the evaluation. All the momenta are of order $gT$. The spectral functions scale as $Im\D_\pm\sim gT$. As a result we obtain $\G_m\sim m^2g^2T^2$. This is to be compared with CS diffusion rate, which scales as $\G_{CS}\sim g^{10}ln g^{-1}T^4$ \cite{Arnold:1996dy,Arnold:1998cy} or $\G_{CS}\sim g^8T^4$ from extrapolation of weak coupling result \cite{Moore:2010jd}. At sufficient weak coupling, the quark mass diffusion rate always dominates the CS diffusion rate.
It is interesting to compare the actual number of the two rates at relevant coupling and mass.
For the former, we quote the strong coupling extrapolation by Moore and Tassler \cite{Moore:2010jd}
\begin{align}
\G_\text{CS}\sim 30\a_s^4T^4.
\end{align}
For the latter, we need to obtain the precise number in $\G_m\sim m^2(gT)^2$ from \eqref{Gm}. We obtain from numerical integration.
\begin{align}\label{Gm_extr}
\G_m\simeq 0.013m^2m_f^2.
\end{align}
We use strange quark mass $m=100MeV$ and use the lattice measured thermal mass \cite{Karsch:2009tp}, which is $m_f\simeq 1.0T$. Taking $T=400MeV$ and $\a_s=0.3$ relevant for heavy ion collisions, we obtain
\begin{align}
\G_\text{CS}\simeq 0.24T^4,\quad \G_m\simeq 0.001T^4.
\end{align}
We found the quark mass effect is much less efficient in axial charge generation. However, the effect of quark mass can be significantly enhanced when temperature approaches transition temperature from above. In this region, the relevant mass parameter is constituent quark mass, which is enhanced by partial chiral symmetry breaking\footnote{We thank Pengfei Zhuang for pointing this out for us. See also related work \cite{Ruggieri:2016asg}}.

Note that in the weakly coupled case, only the generation of axial charge is obtained, the damping effect is not visible. The reason can be understood by making an estimate of damping time scale. Using fluctuation-dissipation theorem, the damping time scale due to quark mass is given by
\begin{align}
\t_m=\frac{\c T}{2\G_m},
\end{align}
where $\c$ is the axial charge susceptibility. To the leading order in $g$, it is given by the free theory result $\c\sim g^0T^2$. We thus obtain $\t_m\sim\frac{T}{m^2g^2}$. Obviously the relaxation is shorter for larger mass, consistent with expectation on general grounds. Note that we assumed $m\sim gT$ in the calculation. The conjugate frequency to this time scale is $q_0\sim g^4T$, which lie well beyond the HTL regime.

\section{Summary}\label{sec_disc}

Let us compare the fluctuation of axial charge in free theory and weakly coupled QGP. First of all, the fluctuations in both cases contain two contributions: one is proportional to susceptibility, originating from charge exchange with heat bath; the other contribution is intrinsic to breaking of axial symmetry. 
Focusing on the contribution from breaking of axial symmetry, we find that unlike the susceptibility term, quantum fluctuation is needed to give a non-vanishing contribution. In case of free theory, the quantum fluctuation is provided by effective Pauli repulsion. This also explains the counter-intuitive result we find: the fluctuation maximizes at zero temperature. It also implies that it is misleading to interpret this contribution as correction to susceptibility. In case of weakly coupled QGP, the fluctuation is given by quark-gluon interaction, which is enhanced by the presence of thermal medium.
The frequency dependence of Wightman correlator of the two cases are given by the following:
\begin{align}\label{summary}
&\lag G^>_G(q_0)\rag^{\text{free}}_G\sim\sqrt{q_0^2-4m^2}\frac{m^2}{|q_0|}\th(|q_0|-2m), \no
&\lag G^>_G(q_0)\rag^{\text{QGP}}_G\sim\frac{m^2m_f^2}{q_0^2}.
\end{align}
In the free case, the Wightman correlator vanishes for $|q_0|<2m$. This gives a flat asymptotic behavior for $\lag N_5^2(t)\rag$ as shown in Fig.~\ref{N5_fig}. In the weakly coupled QGP case, the Wightman correlator is non-vanishing, giving rise to random walk behavior for $\lag N_5^2(t)\rag$ in long time limit. Note that \eqref{summary} is not applicable when $q_0\sim g^2T$. We expect the random walk growth of axial charge to be cut off on an even longer time scale. In order to see the damping effect, we might need kinetic theory to access this time scale \cite{Moore:2006qn}. We leave it for future work.

\section{Acknowledgments}

We are grateful Kenji Fukushima, Lianyi He, Anping Huang, Rob Pisarski and especially Pengfei Zhuang for insightful discussions. SL would like to thank Central China Normal University, Beihang University, Tsinghua University, Peking University, Institute of High Energy Physics for hospitalities where part of this work is done. This work is in part supported by the Ministry of Science and Technology of China (MSTC) under the “973” Project No. 2015CB856904(4) (D. H.), and by NSFC under Grant Nos. 11375070, 11735007, 11521064 (D. H.) and One Thousand Talent Program for Young Scholars (S.L.) and NSFC under Grant Nos 11675274 and 11735007 (S.L.).

\bibliographystyle{unsrt}
\bibliography{Q5ref}

\end{document}